\documentclass[10pt]{article}
\usepackage[explicit]{titlesec}
\usepackage{hyphenat}
\usepackage{ragged2e}
\usepackage[ansinew]{inputenc}
\usepackage{textcomp}
\RaggedRight

\usepackage{garamondx}
\usepackage[T1]{fontenc}
\usepackage{amsmath, amsthm}
\usepackage{graphicx}

\usepackage{soul}
\setul{.6pt}{.4pt}

\usepackage{geometry}
\usepackage{fancyhdr}
\usepackage[labelfont={footnotesize,bf} , textfont=footnotesize]{caption}
\makeatletter
\renewcommand\tagform@[1]{\maketag@@@ {\ignorespaces {\footnotesize{\textbf{Equation}}} #1.\unskip \@@italiccorr }}
\makeatother
\titleformat{\section}
  {\normalfont}{\thesection}{1em}{\MakeUppercase{\textbf{#1}}}
\titlespacing\section{0pt}{0pt}{-10pt}
\titleformat{\subsection}
  {\normalfont}{\thesubsection}{1em}{\textit{#1}}
\titlespacing\subsection{0pt}{0pt}{-8pt}

\makeatletter
\newcommand\sixteen{\@setfontsize\sixteen{17pt}{6}}
\renewcommand{\maketitle}{\bgroup\setlength{\parindent}{0pt}
\begin{flushleft}
\sixteen\bfseries \@title
\medskip
\end{flushleft}
\textit{\@author}
\egroup}
\makeatother

\usepackage[sort&compress]{natbib}
\makeatletter
\renewcommand\@biblabel[1]{\textbf{#1.}\hfill}
\makeatother

\bibpunct{}{}{,~}{s}{,}{,}
\title{Dynamic contrast in scanning microscopic OCT}

\author{Michael M{\"u}nter*$^{1}$, Malte vom Endt$^{2}$, Mario Pieper$^{3,4}$, Malte Casper$^{5}$, Martin Ahrens$^{1,4}$, Tabea Kohlfaerber$^{2}$, Ramtin Rahmanzadeh$^{1}$, Peter K{\"o}nig$^{3,4}$,Gereon H{\"u}ttmann$^{1,2,4}$ and Hinnerk Schulz-Hildebrandt$^{1,2,4}$ \\ \medskip 
$^{1}$Institute of Biomedical Optics, Universit{\"a}t zu L{\"u}beck, 23552 L{\"u}beck, Germany \\ 
$^{2}$Medizinisches Laserzentrum L{\"u}beck GmbH, 23552 L{\"u}beck, Germany \\ 
$^{3}$Institute of Anatomy, Universit{\"a}t zu L{\"u}beck, 23552 L{\"u}beck, Germany \\ 
$^{4}$Airway Research Center North (ARCN), Member of the German Center of Lung Research (DZL), 35392 Gie{\ss}en, Germany \\ 
$^{5}$Laboratory for Functional Optical Imaging, Department of Biomedical Engineering, Columbia University, New York, NY 10027, USA \\ \medskip 
*Corresponding author: mic.muenter@uni-luebeck.de \\ \medskip 
}

\begin{document}
\begin{justify}

\vspace*{.01 in}
\maketitle
\vspace{.12 in}

\section*{abstract}
While optical coherence tomography (OCT) provides a resolution down to 1 \textmu m it has difficulties to visualize cellular structures due to a lack of scattering contrast. By evaluating signal fluctuations, a significant contrast enhancement was demonstrated using time-domain full-field OCT (FF-OCT), which makes cellular and subcellular structures visible. The putative cause of the dynamic OCT signal is ATP-dependent motion of cellular structures in a sub-micrometer range, which provides histology-like contrast. Here we demonstrate dynamic contrast with a scanning frequency-domain OCT (FD-OCT). Given the inherent sectional imaging geometry, scanning FD-OCT provides depth-resolved images across tissue layers, a perspective known from histopathology, much faster and more efficiently than FF-OCT. Both, shorter acquisition times and tomographic depth-sectioning reduce the sensitivity of dynamic contrast for bulk tissue motion artifacts and simplify their correction in post-processing. The implementation of dynamic contrast makes microscopic FD-OCT a promising tool for histological analysis of unstained tissues.

%


\section*{introduction}

OCT is an imaging technique, which is based on the interferometric measurement of backscattered light. Providing fast, high-resolution sectional images of biological tissue, OCT has become a valuable tool in different clinical fields like ophthalmology and dermatology \cite{Drexler2001}. One especially useful characteristic of OCT is the decoupling of the axial resolution and lateral resolution \cite{Fujimoto2000}. For resolving cells and even subcellular structures, it is necessary to increase the resolution of OCT from typically 10 - 15 \textmu m to 1 \textmu m. Optical coherence microscopy (OCM) \cite{Izatt1994}, micro-optical coherence tomography (\textmu OCT) \cite{Liu2011}, and microscopic optical coherence tomography (mOCT) \cite{Pieper2020} typically reach a micron resolution comparable to other microscopic techniques like confocal or nonlinear microscopy \cite{Cox2004}. The use of ultra-broadband light from coupled SLDs, Ti-sapphire lasers or supercontinuum light sources increases systems axial resolution to 1-2 \textmu m. Using a high NA microscope objective, a lateral resolution of 1 \textmu m can be achieved. Loss of lateral resolution outside the focal plane can be compensated by Bessel beams \cite{Leitgeb2006,Curatolo2016}, introducing aberrations \cite{Schulz-Hildebrandt2018} or a numerical correction of the defocus in post-processing \cite{Ralston2007}. In OCT, coherent noise (speckle) reduces contrast \cite{Curatolo2016b,Liba2017} and often makes it impossible to differentiate individual cells based on their scattering properties. Recently, a novel approach, to analyze data of time domain full-field OCT (FF-OCT) was demonstrated and termed as dynamic OCT \cite{Apelian2016}. This method exploits the dynamic scattering changes of metabolically active cellular structures to enhance contrast. FF-OCT illuminates the object over a large area with incoherent light. A horizontal sectional (en-face) image is created by interference between the light scattered by the sample and the light reflected from the reference mirror. To create a tomographic image, it is necessary to move the microscope objective or the sample in axial direction. Dynamic contrast is obtained by analyzing temporal intensity fluctuation in each voxel. Correlation functions or power spectra of the signal fluctuations are evaluated over a few seconds at a typical sampling frequency of 100 Hz \cite{Apelian2016, Thouvenin2017}. This technique detects motion of cellular structures with nanometer sensitivity and 1 \textmu m spatial resolution delivering images which appear similar to conventional histological techniques. Dynamic FF-OCT acquires en-face images of unfixed biological tissue from a depth of up to 100 \textmu m with rich cellular details. Disadvantages of this technique are a low penetration depth in scattering tissues, the limited use of scattered photons from only a 1 \textmu m thick sample plane, the high sensitivity to axial movements and the incapability to quickly acquire large volumes. In contrast, scanned frequency-domain OCT (FD-OCT) provides real tomographic imaging simultaneous over a certain depth range and uses all photons. Additionally, it has more degrees of freedom in defining the imaged region. However scanned OCT typically lacks lateral phase stability and suffers from a shallow depth of focus, when using high-NA objective lenses, e.g. at 1 \textmu m focus diameter the confocal parameter is only 17 \textmu m. Outside this range, lateral resolution decreases and dynamic contrast may be annihilated. This work demonstrates that scanned FD-OCT produces surprisingly rich dynamic contrasted B-scans and volumetric images of freshly excised murine tissue samples - tongue and liver, despite high NA.
\section*{methods and procedures}
\vspace{0.25 cm} 
\begin{figure}[h!]
\centering
\includegraphics[width=0.9\textwidth]{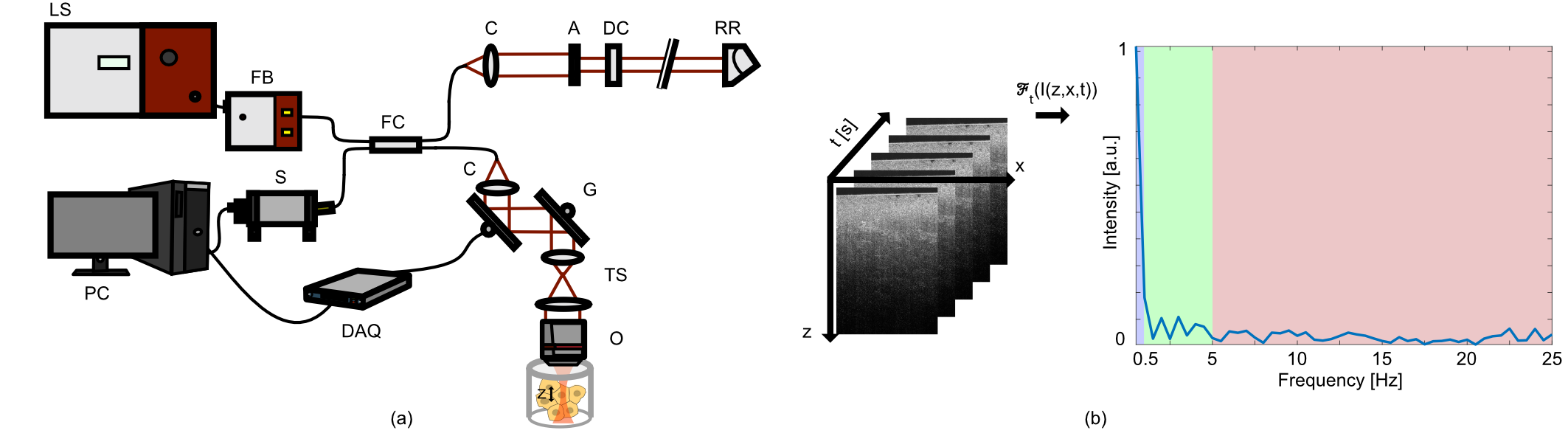}
\caption{(a) Schematic of the mOCT setup. LS: light source, FB: filter box, FC: 50/50 fiber coupler, C: collimators, G: xy galvanometer mirror scanner, TS: scan lenses, O: microscope objective, DC: dispersion compensation, A: aperture, RR: retroreflector, DAQ: data acquisition device, S: spectrometer; PC: computer for data acquisition and scanning control. Nanometer cellular motion is detected by intensity fluctuations (b) The dOCT image is generated by t color-coding of three frequency bands after Fourier transforming the temporal signal fluctuations in each pixel of the B-scan image stack}\label{Fig1}
\end{figure}\vspace{0.5 cm} 

The mOCT setup illustrated in Fig. \ref{Fig1}, uses a supercontinuum light source (SuperK EXTREME-EXW-OCT, NKT Holding, Denmark) in combination with spectrometer covering a spectral range from 550 to 950 nm in order to achieve 1 \textmu m axial resolution. Emission of the supercontinuum light source was bandpass filtered (SuperK SPLIT, NKT Holding, Denmark) and injected to a fiber-based Michelson interferometer. A broadband 50/50 fiber coupler (TW630R5A2, Thorlabs Inc., U.S.) splits the light into the sample and reference arm. Light in the sample arm is collimated (60FC-L-4-M25-02, Sch{\"a}fter + Kirchhoff GmbH, Germany) and scanned via an x/y galvanometer scanner module (6210H, Cambridge Technology, U.S.) and an achromatic telescope (SL50-CLS2 and TL200-CLS2, Thorlabs Inc., U.S.) to the back focal plane of a 10x/0.3 NA microscope objective (HCX APO L 10x/0.3 WUVI, Leica Microsystems, Germany). Radiant flux at the sample was 40 mW. Light in the reference arm was collimated (60FC-L-4-M25-02, Sch{\"a}fter + Kirchhoff GmbH, Germany), attenuated by a variable neutral density filter (NDC-50C-2M, Thorlabs Inc., U.S.), and matched in dispersion to the light of the sample arm by a 14 mm thick piece of SF57 glass substrate (Casix Inc., China). Reference light from a retroreflector (PS975M-B, Thorlabs Inc., U.S.) and sample light were brought to interference in the custom-designed spectrometer (Thorlabs GmbH, Germany). Two different cameras (SprintSpL4096 140km, Basler, Germany and OctoPlus CL, Teledyne, e2v, Canada) were used for achieving A-scan rates of up to 127 kHz and 248 kHz, respectively. Synchronization of the scanner, spectrometer and data acquisition software was realized using a USB data acquisition device (NI USB-6251, National Instruments, U.S.). For volumetric imaging, at each line 100 B-scans with 500 A-scans each were taken at 30 kHz A-scan rate with the Sprint SpL4096 camera, which results in an effective B-scan rate of 46 Hz. Adjacent A-Scans were axial aligned using the reflection of the glass plate's surface to compensate for sample motion and image field curvature. Volumes were recorded by stacking a series of B-scans in y-direction.  With the OctoPlus CL 150 B-scans with 500 A-scan, each were sampled at 100 kHz A-scan rate. With an effective B-scan rate of 108 Hz and a total recording time of 1.39 s, the frequency of signal fluctuations was measured between 0 to 54 Hz. Raw data from the spectrometer was Hann-windowed and Fourier transformed to obtain the OCT images. Residual dispersion was numerically compensated using a 5th order polynomial for the spectral phase error correction. Polynomial coefficients were determined by optimizing image quality, which was measured by the Shannon entropy \cite{Wojtkowski2004, Schulz-Hildebrandt2017}. Spectral processing and dynamic contrasting of the data were performed in MATLAB (MATLAB R2019b, The MathWorks, Inc., U.S.). At each voxel, the temporal variations of the absolute value of the OCT signal was evaluated. Similar to the data evaluation in dynamic FF-OCT, the time series was Fourier transformed and the integral amplitude was calculated in three frequency bands (Fig. 1b). For each pixel the three resulting values were color-coded in an RGB image, representing different time scales of motion activity. Blue represents slow motion frequencies (0-0.5 Hz), green medium motion frequencies (0.5-5 Hz), and red fast motion (5-25 Hz). For the representation of the images, the color channels were scaled logarithmically. Image contrast was enhanced using contrast-limited adaptive histogram equalization \cite{Zuiderveld1994}. Brightness outside the focus was adjusted to the peak intensity in the focus. Finally, a 3x3 median filter was applied to each channel. Reference measurements were acquired with a commercially available FF-OCT device (LightCT, LLTech Inc., France), which was designed for real-time optical biopsy using dynamic OCT contrast. LightCT uses a Linnik interferometer and a temporally and spatially incoherent light source \cite{Manen2017}. Using a spectral width of 100 nm at a central wavelength of 565 nm and a 0.3 NA water immersion objective, a nearly 1 \textmu m isotropic resolution is achieved. Our frequency bands for color-coding were chosen directly to match those of the LightCT. For immobilizing of the sample, freshly excised tongue and liver tissue of C57BL/6 mice were placed in Ringer's solution in the specially designed sample holder of the LightCT. Tissues were imaged by slightly pressing the tissue surface against the quartz cover glass plate to which the objective was coupled using silicon immersion oil. 

\section*{results}
\vspace{0.25 cm} 
\begin{figure}[h]
\centering
\includegraphics[width=1\textwidth]{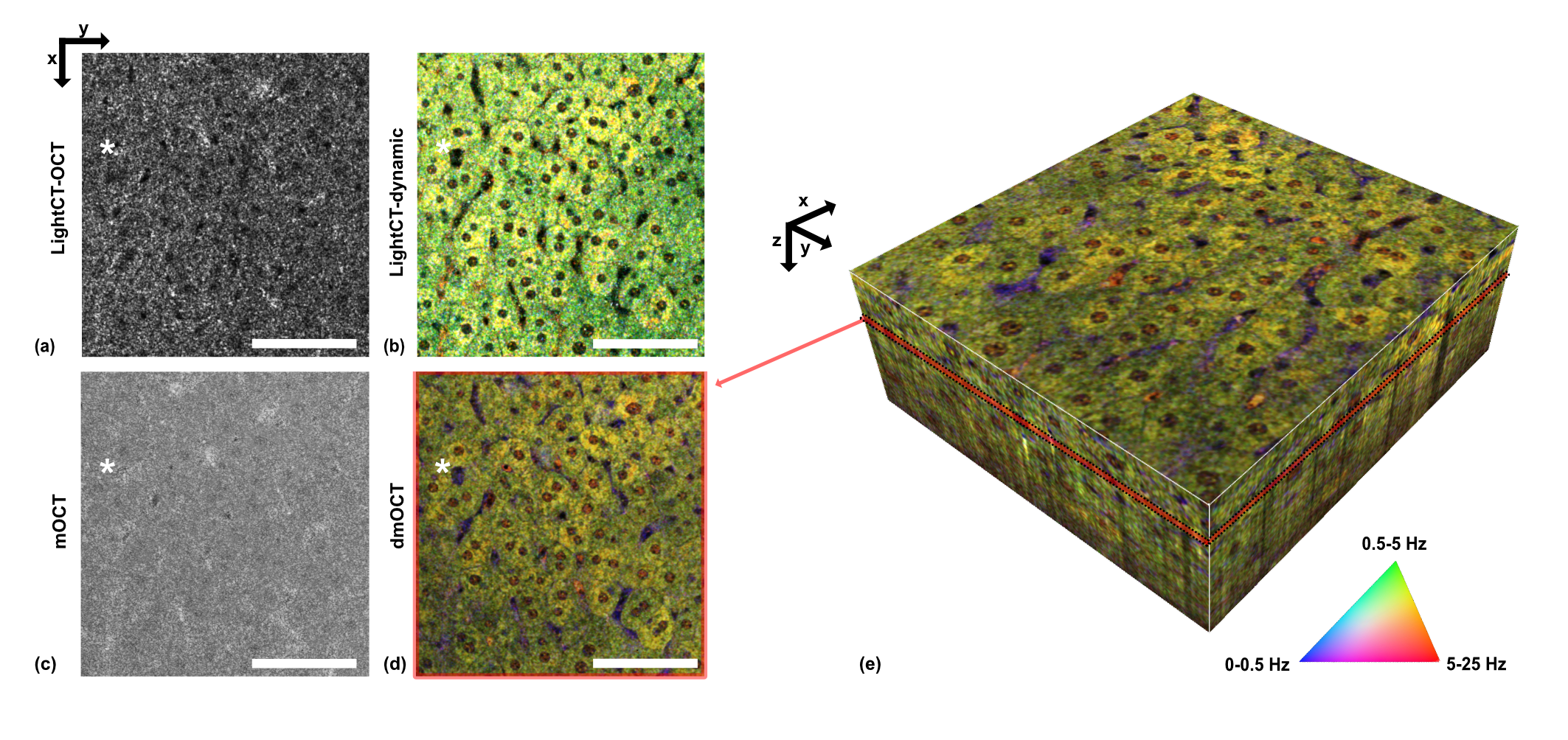}
\caption{(a) LightCT en-face OCT image of murine liver; (b) LightCT dynamic OCT en-face image with corresponding regions marked with (*) (c) Averaged en-face image from the scanned mOCT, which was acquired at the same region; (d) in the corresponding dynamic mOCT image hepatocytes become visible with nuclei (e) Cropped volume representation; size: 135x270x285 \textmu m (zxy) (Visualization available on request);  scalebar: 100  \textmu m}\label{Fig2}
\end{figure}\vspace{0.5 cm} 
Despite the sufficient resolution, mOCT and LightCT OCT images of liver tissue showed only weak contrast. The murine liver shown in Fig. \ref{Fig2} was imaged over a FOV of 550x270x285 \textmu m (zxy). Fine cellular details of the tissue are difficult to identify without dynamic contrasting in averaged en-face OCT image planes shown in Fig. \ref{Fig2}a and c. Brighter details might indicate extracellular matrix structures and spherical voids cell nuclei. However, images with the dynamic contrast clearly show cellular structures (Fig. \ref{Fig2}b and d). Cell nuclei and cell borders can be distinguished with high contrast. Cell nuclei appear with increased movement activity in red while the extracellular matrix in blue indicating slow movement. Dynamic mOCT and LightCT images depict the tissue structures in similar color, however, intensity and composition of color/motion components are slightly different. A volumetric mOCT image cropped in z recorded is given in Fig. \ref{Fig2}e and Visualization 1 (available on request). 
Similar results were achieved imaging the bottom of murine tongue. The murine tongue shown in Fig. \ref{Fig3} was imaged over a FOV of 660x500 \textmu m (zx). This part of the tongue consists of a cornified stratified squamous epithelium and a subepithelial layer of connective tissue followed by skeletal muscle cells (Fig. \ref{Fig3}a). The average of 150 mOCT B-Scans does only shows a faint image of the tissue structures (Fig. 3b). On the right side of the OCT image four tissue layers can only be guessed. On the left side, a layered structure is not visible. Using dynamic contrast, histologically relevant layers become clearly visible in the whole field of view (Fig. \ref{Fig3}c). Below the glass plate, the cornified layer, the layer of the squamous epithelium with the transition to the granular and spinous layer is contrasted as a purple and a green-purple layer, respectively. Deeper, the basal layer, which is characterized by yellow-colored cell nuclei, is displayed. Since the optical focus was placed in this region the basal layer is imaged at the highest resolution and structures even within the nuclei became discernible. Below, the lamina propria is visible as a purple band again followed by the green-colored skeletal muscle fibers. Shifting the focus below the basal layer increases the visibility of the muscle layer (Fig. \ref{Fig3}d). Dynamic contrast mOCT images show an excellent match to structures in HE stained histological sections (Fig. \ref{Fig3}a, c-d). The dynamic contrast images are completely free of speckle noise and have remarkably high contrast. Cellular structures are discernible even well beyond the focal plane, which are typically invisible in averaged OCT images even when exactly focused. The high contrast permits a reliable segmentation of layers and even individual basal cells.
\vspace{0.25 cm} 
\begin{figure}[h]
\centering
\includegraphics[width=1\textwidth]{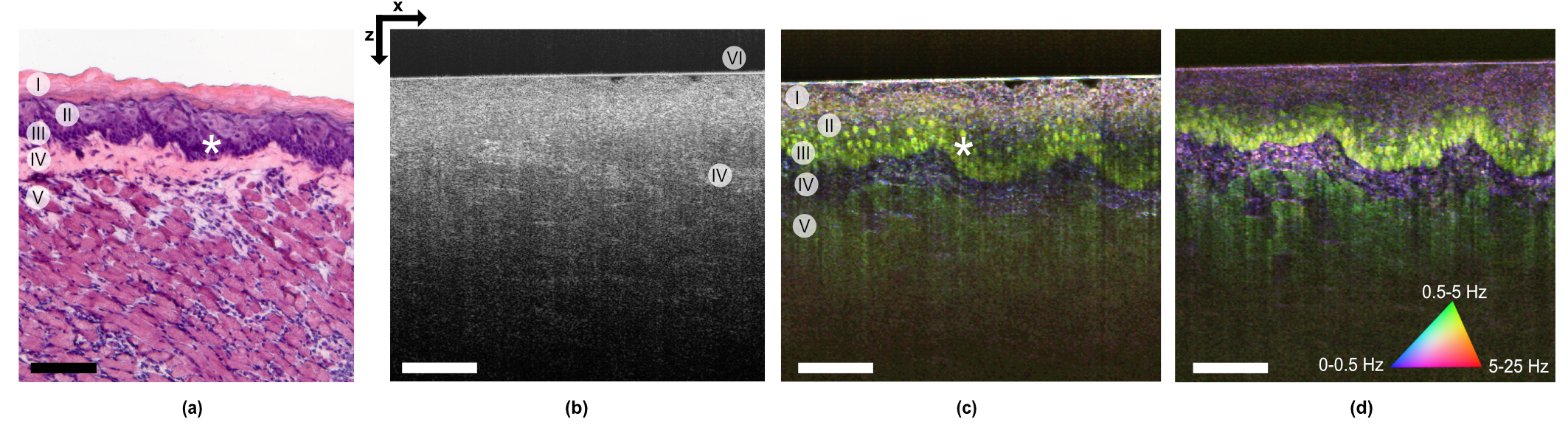}
\caption{HE stained histology of the imaged sample at a different location (I) Cornified layer, (II) Granular \&   spinous layers, (III) Basal layer, (IV) Lamina propria, (V) muscle, (VI) Glass plate (b) OCT image of mouse tongue; lamina propria (IV) can be identified by brighter contrast (c) Corresponding dynamic contrast mOCT image with focus in basal cell layer; (I-V) and even cell nuclei (*) are visible (d) Dynamic contrast mOCT image  with focus in the lamina propria; cropped image with size of 460x500 \textmu m (zx); scalebar: 100  \textmu m}\label{Fig3}
\end{figure}
\vspace{0.5 cm} 
\section*{discussion}
These results demonstrate, that implementing dynamic contrast in a scanning mOCT is possible and yields dramatically increased contrast that allows identification of individual epithelial cells. This is somewhat surprising as dynamic contrast is inferred from motion on nanometer scale and scanning can lead to additional phase noise potentially overcasting phase fluctuation caused in the sample. While the concept of dynamic contrast was developed for FF-OCT, here we show, it can also be used with scanning mOCT. This offers several advantages. First, it provides inherently sectional images across the tissue layers, which show several tissue layers at a glance as known from conventional histological sections that are routinely used in pathology. Reconstruction of a B-scan from a series of en-face FF-OCT images is time-consuming and image quality can easily be degraded by small sample motions. Second, acquiring a sectional image plane reduces the sensitivity to axial sample motion, which has the most severe influence on dynamic contrast. While in FF-OCT systems axial sample motion is difficult to detect and correct \cite{Scholler2020}, in sectional scans each A-scan can be individually corrected by multiplying an appropriate phase factor. Third, the parallel measurement of the complete depth information in FD-OCT potentially increases speed for volumetric imaging. Dynamic contrast OCT needs to measure each pixel over a few seconds with a temporal resolution of about 10 ms. This limits the imaging speed. FF-OCT, which currently uses a 2 Megapixel camera is able to image about 1 million volume elements per second. At 1 \textmu m resolution this converts into an imaging time of 1 s/mm2 for one en-face image. For volumetric imaging over 300 \textmu m depth, time is increased to 300 s or 5 min per mm2. Scanned OCT with B-scan based evaluation of the signal fluctuation does not increase imaging speed for volume acquisition. However, FD-OCT with MHz A-scan rate could acquire a whole volume within 10 ms \cite{Wieser2010} and complete volumetric dynamic contrast OCT imaging within a few seconds. Currently, our imaging speed is limited by three components – galvo scanner, line scan camera and relative intensity noise (RIN) of supercontinuum light sources. No negative effects of RIN on image quality could be found yet and recently we were able to demonstrate mOCT imaging at 1 \textmu m isotropic resolution with 600 kHz A-scan rate and nearly 12 volumes per second \cite{Muenter2019}. However, fast volumetric scanning is still a challenge. Linear, single direction scanning introduces considerably overhead, which can only be compensated by bidirectional or resonant scanning. The influence of fast scanning on dynamic contrast is currently under investigation. Forth, dynamic contrast with scanning OCT is easier to implement in an endoscopic system. Since technologies using scanning OCT imaging in rigid and flexible endoscopes have already been developed \cite{Schulz-Hildebrandt2018b}, dynamic contrasting, as an imaging routine, could equip such devices with the additional contrast needed to facilitate non-invasive, \textit{in vivo} optical histology. In conclusion, the analysis of the signal dynamics of cellular resolution scanned FD-OCT visualizes epithelial tissues comparable to conventional histology without the need for tissue processing. We found that scanning mirrors do not interfere with the dynamic signal extraction. By evaluating the fluctuation of the OCT signal, speckle-free B-scan and volumetric imaging with high contrast and resolution is achieved. Spectral analysis discriminates and visualizes important cellular components like nuclei and cytoplasm and allows the identification of different tissue components such as epithelium, connective tissue and muscle. We expect that scanning FD-OCT is more robust against sample motion than FF-OCT and can increase imaging speed by two orders of magnitude. Given these assets, dynamic contrast OCT might even fulfill the 20 years old promise of optical biopsies by OCT \cite{Fujimoto2000}.

\section*{Funding}
This research was funded by the European Union project within Interreg Deutschland-Denmark from the European Regional Development Fund (ERDF) in the project CELLTOM and German Ministry of Research, Innovation and Science, Helmholtz Center Munich of Health and Environment DZL-ARCN (82DZL001A2) and German Research Foundation (DFG, RA 1771/4-1, EXC 2167).
\end{justify}

\end{document}